\documentclass{aa}
\usepackage{txfonts}
\usepackage{graphicx}
\usepackage{natbib}
\bibpunct{(}{)}{;}{a}{}{,}

\begin{document}

\title{Is Praesepe really different from the coeval Hyades cluster? \\
The XMM-{\it Newton} view}

\author{E. Franciosini\inst{1} \and S. Randich\inst{2} \and R.
Pallavicini\inst{1}}

\institute{
INAF/Osservatorio Astronomico di Palermo, Piazza del Parlamento 1, I-90134
Palermo, Italy
\and
INAF/Osservatorio Astrofisico di Arcetri, Largo E. Fermi 5, I-50125 Firenze,
Italy
}

\offprints{E. Franciosini, \email{francio@astropa.unipa.it}}

\date{Received  / Accepted  }

\abstract{
The Praesepe open cluster represents a puzzle since it has about the same
age as the Hyades, and only slightly different metallicity, yet previous
{\it ROSAT} observations resulted in a detection rate of cluster sources
significantly lower than for the Hyades. We present a new 50 ksec
observation of Praesepe performed with the EPIC instrument on board XMM-{\it
Newton}, which resulted in the detection of $\sim 200$ sources, including 48
cluster members. We detected all solar-type (F--G) stars in the field of
view, $\sim 90$\% of the K stars and $\sim 70$\% of the M stars. We find
that the distribution of X-ray luminosities of solar-type Praesepe members
is comparable to that of the Hyades, in contrast with the previous {\it
ROSAT} results; however, the disagreement between the {\it ROSAT} and
XMM-{\it Newton} results appears to be mostly due to X-ray faint Praesepe
members falling outside the XMM-{\it Newton} field of view, while it is
considerably reduced when considering only the subsample of stars in the
{\it ROSAT} survey in common with the present observation. The finding
supports an earlier suggestion that Praesepe may be formed by two merged
clusters of different age.
\keywords{Open clusters and associations: individual: Praesepe -- stars:
activity -- stars: coronae -- X-ray: stars}
}

\titlerunning{Is Praesepe different from the Hyades? The XMM-{\it Newton}
view}

\maketitle

\section{Introduction}

The discrepancy in the average X-ray properties of the \object{Hyades} and
\object{Praesepe} clusters represents one of the most surprising results in
the context of {\it ROSAT} observations of open clusters. The two clusters
have a very similar age (about 600~Myr); hence, based on the general
age--activity relation which was shown to hold by early {\it Einstein}
observations and confirmed by {\it ROSAT} \citep[e.g.][and references
therein]{randich00}, one would have expected that they share very similar
X-ray properties. On the contrary, \citet[hereafter RS95]{randsch95} found
that the bulk of Praesepe population is significantly underluminous in
X-rays with respect to the Hyades. More specifically, RS95 carried out a
{\it ROSAT} PSPC raster scan covering a large area of the cluster ($\approx
4\degr \times 4\degr$) and detected only about 30\% of the late-F and G-type
stars, above a sensitivity threshold ranging between $\sim 2\times 10^{28}$
and $1 \times 10^{29}$~erg~s$^{-1}$. This result must be compared with the
$\sim 90$\% detection rate among the Hyades members with the same spectral
type \citep{stern95} and with a median luminosity for this cluster $L_{\rm
X} \simeq 8\times 10^{28}$~erg~s$^{-1}$. Later-type stars in Praesepe were
characterized by even lower detection rates ($\sim 10$\%), again in contrast
with the results for the Hyades \citep{pye94,stern95}. As a consequence, the
X-ray luminosity distribution functions (XLDFs) of Praesepe stars of
different spectral types, and in particular the XLDF of solar-type stars,
lied considerably below those of the Hyades and were dominated by upper
limits. We mention that a large fraction of the Praesepe {\it ROSAT} survey
had a similar sensitivity to the {\it ROSAT} observations of the Hyades;
thus the hypothesis that the difference in the XLDFs of the two clusters
could be due to different sensitivities appeared rather unlikely. 

As pointed out by RS95, {\it ROSAT} observations of Praesepe cast doubts on
the universality of the activity--age relation; in other words, this finding
may imply that the X-ray activity vs. age relationship is not unique and
that a given cluster is not necessarily representative of the whole stellar
population at that age.

In order to understand the anomalous X-ray emission of Praesepe members and,
in particular, to ascertain whether the available optical catalogs of
Praesepe were contaminated by a significant number of non-members which
could bias the X-ray results, \citet{barrado98} carried out optical
follow-up spectroscopic observations of a sample of F-K stars and M dwarfs
in Praesepe. Based on the derived radial velocities, they could exclude that
a significant number of non-members was present in the Praesepe catalog used
for the X-ray analysis. In addition, H$\alpha$ was measured for the M dwarf
sample; a statistical comparison showed that for these stars the
distribution of H$\alpha$ chromospheric emission is similar to the Hyades,
at variance with the X-ray results. Finally, Barrado y Navascu\`es et al.
also analyzed a few {\it ROSAT} PSPC pointings retrieved from the archive
and covering a fraction of the raster scan; based on these X-ray
observations, they confirmed that the discrepancy in the X-ray properties of
the Hyades and Praesepe could not be explained by an inadequate sensitivity
of the raster scan data. Similarly to RS95, they concluded that the
difference was most likely real.

Various hypotheses were proposed to explain the discrepancy between the two
clusters. Namely, 1. a difference in the rotation rate distributions, 2. a
difference in metallicity, and 3. the possibility that Praesepe may consist
of two merged clusters. As it is now well known, the main parameter
determining the level of X-ray emission at a given mass (or spectral type)
is rotation and the X-ray vs. age relationship is indeed an
X-ray--rotation--age relation \citep[e.g.][]{jeffries99}. If the rotation
rates of the bulk of Praesepe population were lower than the Hyades, this
could explain the X-ray results. Rotational velocities and/or periods for
late-type Praesepe single members have never been published in a tabular
form; however, projected rotational velocities for Praesepe stars with $B-V
> 0.6$ are available and have been published by \citet{merm97ms} as a
figure. The distributions of projected rotational velocities $v\sin i$ of
the Hyades and Praesepe appear very similar; as stressed by Mermilliod, the
comparison between the two distributions does not therefore support the idea
that the discrepancy in the X-ray properties is due to a difference in the
cluster rotation rate distributions. 

For Praesepe an almost solar metallicity has been determined
($\textrm{[Fe/H]} =0.038 \pm 0.039$) by \citet{friel92}, while the same
authors derived $\textrm{[Fe/H]}=0.127 \pm 0.022$ for the Hyades. This
difference in metallicity implies a difference in the depth of the outer
convective zone and, in principle, in the dynamo efficiency. In a recent
study, \citet{pizzol01} theoretically investigated the dependence of X-ray
emission on metallicity for late-F, G and K dwarfs, however their results
for late-F and G stars remain inconclusive. In fact, they found that for
late-F and G stars the convective turnover time, which is a key parameter in
determining the level of X-ray emission, decreases with increasing
metallicity; at the same time the coronal plasma is predicted to radiate
more efficiently for higher metallicities and the two effects almost
compensate. Since their relative contribution depends on the exact stellar
color and plasma temperature, \citet{pizzol01} could not draw any definitive
conclusion on the metallicity dependence of X-ray emission of late-F and
G-type stars. Thus, the possibility that the discrepancy in the X-ray
properties of the two clusters may be due at least in part to the different
[Fe/H] contents remains open.

Finally, in a recent work \citet{holland00} hypothesized the presence of a
subcluster lying about 3~pc away from the center of Praesepe. They pointed
out that the brightest X-ray sources were found almost exclusively in the
main cluster, and suggested that there could be a difference in age between
the main cluster and the subcluster.

Praesepe is not the only exception to the age--X-ray activity relation;
other possible X-ray underluminous clusters challenging the
age--rotation--activity relationship were found, such as NGC~3532
\citep{francio00}, NGC~6633 \citep{totten00,francio00asp,harmer01} and
Stock~2 \citep{sciorti00}. The results for these clusters however, contrary
to the Praesepe case, may be biased due to the not deep enough X-ray
surveys, and/or to the incompleteness of optical catalogs and contamination
by non-members, or to uncertainties in the cluster distances; additional
optical and X-ray observations are needed to confirm the reality of these
possible exceptions.

In order to further investigate the anomalous X-ray properties of Praesepe
and to better constrain the XLDF of its members by detecting a larger
fraction of cluster members, we carried out a deep XMM-{\it Newton} pointing
of the cluster, covering a field of $\approx 30\arcmin \times 30\arcmin$ in
the central regions of the cluster. The nominal pointing direction is
$\alpha = 08^{\rm h} 39^{\rm m} 58^{\rm s}$, $\delta = 19\degr 32\arcmin
29\arcsec$ (J2000).

Our paper is organized as follows. In the next section we describe the input
optical catalog; observations and data analysis are described in
Sect.~\ref{obs}. In Sect.~\ref{results} the results of our analysis and the
comparison with the Hyades cluster are presented and discussed; conclusions
are given in Sect.~\ref{concl}.

\section{The optical catalog}
\label{opt}

We have constructed a catalog of probable and possible members of the
Praesepe cluster using as a basis the compilation by RS95 from the proper
motion surveys by \citet{kw27} and \citet{jc83}, and from the photometric
and proper motion study by \citet{js91} (see RS95 for details). We have
updated this catalog using the deep proper motion and photometric survey by
\citet{hshj95}, who covered a field of $\sim 19$ deg$^2$ down to $R \sim 20$
and $I\sim 18.2$, providing data for 515 stars with membership probability
greater than 40\%, and the proper motion study by \citet{wang95}, who
provide data for 924 stars in a $90\arcmin \times 90\arcmin$ region. We also
added stars from the radial velocity study by \citet{merm90} and from the
photometric surveys of low-mass stars and brown dwarf candidates by
\citet{wrs95} and \citet{pinfield97}. Additional photometry and radial
velocity data, as well as information on binarity, was retrieved from
several studies. We selected as probable or possible members those stars
with radial velocity within 5 km/s of the cluster mean $v_{\rm r}$ (34.5
km/s), when available, or with membership probability from proper motions
greater than 75\%, and having photometry consistent with cluster membership,
i.e. falling within 0.2 mag below and 0.75 mag above the cluster main
sequence. Combining all the available membership information, we assigned to
each star a total membership flag (Y for probable members and Y? for
possible members). For stars with no proper motion and radial velocity data,
the final membership is based on photometry only.

Recently, \citet{adams02} performed a new proper motion and spectroscopic
study based on data from the 2MASS catalog covering a region of 100 deg$^2$.
We added to our catalog those stars that were not already included in the
list. The membership probabilities derived by \citet{adams02} are lower than
in previous studies due to a stronger contamination by field stars;
following these authors, we accepted as candidate members stars with $P \ge
20$\%.

The resulting catalog contains 150 stars falling in the XMM-{\it Newton}
field of view, of which 61 are probable or possible members, including two
of the five Praesepe giants, and two candidate members from \citet{adams02}.

\begin{figure}
\resizebox{\hsize}{!}{\includegraphics[clip]{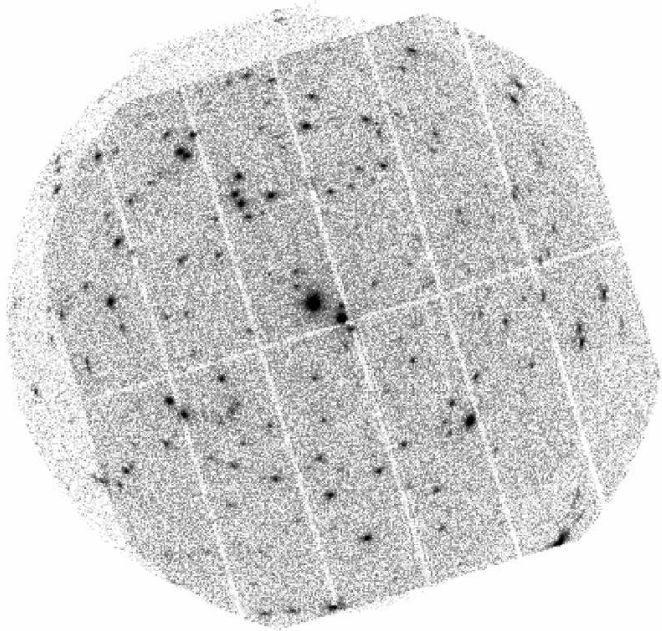}}
\caption{Composite {\it EPIC} (MOS1+MOS2+PN) image of the Praesepe field}
\label{image}
\end{figure}

\section{Observations and data analysis}
\label{obs}

The Praesepe cluster was observed as part of the Guaranteed Time programme
using the EPIC cameras on board XMM-{\it Newton}. The observation (ID
0101440401) started at 21:33 UT on November 7, 2000 and ended at 10:11 UT on
November 8, 2000, for a total exposure time of 45.5 ksec in the two MOS
cameras and 43 ksec in the PN camera. All cameras were operated in Full
Frame mode; the thick and the medium filters were used for MOS and PN,
respectively.

Data analysis was carried out using the standard tasks in SAS v.5.3.3.
Calibrated and cleaned event files were derived from the raw data using the
standard pipeline tasks {\sc emchain} and {\sc epchain} and then applying
the appropriate filters to eliminate noise and bad events. The PN data were
affected by a short period of high background due to proton flares; we have
therefore filtered the event file in order to eliminate the affected time
segment. The resulting effective exposure time in the PN is 41 ksec. We have
limited our analysis to the $0.3-7.8$ keV energy band, since data with
energy below 0.3 keV is mostly unrelated to bona-fide X-rays, while above
7.8 keV only background is present. Images were accumulated from the event
files using a binsize of $4\arcsec$.

\subsection{Source detection}
\label{detect}

Source detection was first performed on the three individual images using
the Maximum Likelihood (ML) algorithm. After checking the relative alignment
of the three EPIC cameras by comparing the positions of common sources, we
merged the three event files into a single dataset; since the median
relative shifts between the three instruments are $\la 2\arcsec$, i.e. less
than one image pixel, we did not apply any correction to the data. The
combined EPIC image is shown in Fig.~\ref{image}.

A combined exposure map was obtained by summing the individual exposure maps
of the single instruments with an appropriate scaling factor for PN, in
order to take into account the different sensitivities of MOS and PN. The
scaling factor was derived by measuring the ratio of PN to MOS count rates
for the detected sources with a stellar counterpart: we found a median ratio
PN/MOS $\simeq 4.6$ in the 0.3-7.8 keV energy band. This implies for the
merged dataset an equivalent MOS exposure time of $\sim 280$ ksec.

\begin{table*}[!ht]
\caption{X-ray detected Praesepe members. $\Delta r$ is the distance to the
optical position. The column labeled ``set'' indicates whether the source
was detected on the summed dataset (S) or only on the MOS (M) or PN (P)
single datasets. Count rates are MOS equivalent count rates, except where
set = P, when they are PN count rates. Luminosities were derived using the
appropriate conversion factors (see text)}
\begin{tabular}{lcccrcrrrlll}
\hline\hline
\noalign{\smallskip}
Name& RA$_{\rm X}$& DEC$_{\rm X}$& $\Delta r$& ML& set& count rate& 
$L_{\rm X}$ &$V$& $B-V$& Mem& Notes \\
& \multicolumn{2}{c}{(J2000)}& ($\>\arcsec\>$)& & & (cts/ksec)& ($a$)& &
$(R-I)_{\rm c}$\hskip -3pt *& & ($b$)\\
\hline
\noalign{\smallskip}
KW 154  & $8\>39\>09.24$& $19\>35\>32.5$& 2.02&    39.5& S& 
  $ 0.71 \pm 0.14$&   1.87&  8.50& 0.25 & Y &         \\
KW 155  & $8\>39\>10.20$& $19\>40\>45.1$& 2.73&    24.8& S& 
  $ 0.70 \pm 0.17$&   1.87&  9.42& 0.41 & Y &         \\
KW 181  & $8\>39\>25.05$& $19\>27\>35.2$& 1.64&  8056.9& S& 
  $17.45 \pm 0.43$&  46.28& 10.47& 0.59 & Y & SB, X   \\
KW 184  & $8\>39\>28.66$& $19\>28\>26.5$& 1.56&   280.2& S& 
  $ 1.68 \pm 0.16$&   4.45& 11.56& 0.90 & Y & SB2, X  \\
KW 198  & $8\>39\>38.42$& $19\>26\>28.5$& 1.57&   229.5& S& 
  $ 1.33 \pm 0.13$&   3.53& 12.59& 0.96 & Y &         \\
KW 208  & $8\>39\>46.00$& $19\>22\>01.5$& 3.22&  1139.7& S& 
  $ 8.06 \pm 0.44$&  21.38& 10.66& 0.58 & Y &         \\
KW 212  & $8\>39\>50.76$& $19\>32\>27.2$& 0.68& 15189.1& S& 
  $17.37 \pm 0.33$&  46.08&  6.59& 0.96 & Y & K0III, X\\
KW 213  & $8\>39\>50.91$& $19\>33\>02.9$& 1.10&   212.0& S& 
  $ 1.20 \pm 0.12$&   3.18& 11.81& 0.80 & Y & X       \\
KW 217  & $8\>39\>52.45$& $19\>18\>45.7$& 1.83&  1399.0& S& 
  $ 9.13 \pm 0.49$&  24.22& 10.23& 0.51 & Y & X       \\
KW 224  & $8\>39\>56.54$& $19\>33\>11.3$& 0.98& 53890.7& S& 
  $45.14 \pm 0.50$& 119.73&  7.32& 0.19 & Y & SB1, X  \\
KW 236  & $8\>39\>59.99$& $19\>33\>59.9$& 1.50&   145.5& S& 
  $ 0.74 \pm 0.09$&   1.96& 11.94& 1.00 & Y & SB, X   \\
KW 237  & $8\>40\>00.05$& $19\>34\>41.8$& 1.48&    99.2& S& 
  $ 0.63 \pm 0.09$&   1.67& 12.86& 1.00 & Y & X       \\
KW 246  & $8\>40\>03.95$& $19\>47\>06.2$& 4.42&    17.9& M& 
  $ 3.49 \pm 1.05$&   9.25& 12.01& 0.84 & Y?& SB1     \\
KW 250  & $8\>40\>04.94$& $19\>43\>46.2$& 0.78&   173.0& S& 
  $ 1.57 \pm 0.17$&   4.16&  9.79& 0.47 & Y &         \\
KW 257  & $8\>40\>06.30$& $19\>18\>27.1$& 1.70&   211.9& S& 
  $ 3.27 \pm 0.35$&   8.67& 11.00& 0.77 & Y & SB, PhB \\
KW 263  & $8\>40\>09.68$& $19\>37\>18.3$& 1.51&   348.1& S& 
  $ 1.61 \pm 0.13$&   4.28& 12.02& 0.81 & Y & X       \\
KW 268  & $8\>40\>12.35$& $19\>38\>22.3$& 0.90&  1191.2& S& 
  $ 3.70 \pm 0.19$&   9.83&  9.89& 0.48 & Y & SB1     \\
KW 279  & $8\>40\>20.72$& $19\>41\>12.3$& 0.37&   525.7& S& 
  $ 3.52 \pm 0.26$&   9.35&  7.70& 0.20 & Y & SB1     \\
KW 283  & $8\>40\>22.13$& $19\>40\>11.5$& 0.66&  1335.6& S& 
  $ 5.18 \pm 0.24$&  13.74&  6.44& 1.02 & Y & K0.5IIIa\\
KW 287  & $8\>40\>23.28$& $19\>40\>22.8$& 1.18&  2369.4& S& 
  $ 7.98 \pm 0.31$&  21.16& 10.37& 0.59 & Y & SB1     \\
KW 288  & $8\>40\>22.75$& $19\>27\>54.0$& 0.52&  1863.4& S& 
  $ 7.67 \pm 0.33$&  20.35& 10.71& 0.58 & Y & X       \\
KW 293  & $8\>40\>25.60$& $19\>28\>33.7$& 1.71&  1609.4& S& 
  $ 4.70 \pm 0.21$&  12.47&  9.89& 0.48 & Y & X       \\
KW 295  & $8\>40\>26.12$& $19\>41\>11.6$& 0.42&   227.1& S& 
  $ 1.84 \pm 0.18$&   4.87&  9.41& 0.41 & Y &         \\
KW 299  & $8\>40\>27.50$& $19\>39\>21.9$& 2.39&    45.1& S& 
  $ 0.54 \pm 0.10$&   1.44& 13.17& 1.07 & Y &         \\
KW 313  & $8\>40\>33.37$& $19\>38\>02.7$& 2.75&   196.5& S& 
  $ 1.71 \pm 0.19$&   4.54& 12.20& 0.89 & Y &         \\
KW 322  & $8\>40\>39.92$& $19\>40\>10.8$& 4.01&   685.4& S& 
  $ 4.15 \pm 0.28$&  11.01& 10.87& 0.68 & Y & SB1, X  \\
KW 326  & $8\>40\>42.40$& $19\>33\>59.7$& 2.86&   133.1& S& 
  $ 1.21 \pm 0.15$&   3.20& 11.34& 0.72 & Y &         \\
KW 334  & $8\>40\>47.97$& $19\>39\>32.3$& 1.32&   125.5& S& 
  $ 3.39 \pm 0.47$&   8.99& 11.02& 0.72 & Y & PhB, X  \\
KW 341  & $8\>40\>52.40$& $19\>28\>58.0$& 2.76&   234.8& S& 
  $ 4.78 \pm 0.52$&  12.66& 10.30& 0.52 & Y & SB1     \\
JC 143  & $8\>39\>03.01$& $19\>31\>58.5$& 1.57&   377.2& S& 
  $ 3.73 \pm 0.31$&   9.90& 16.92& 1.57 & Y & PhB, X  \\
JS 267  & $8\>39\>04.09$& $19\>31\>24.5$& 3.02&    26.9& S& 
  $ 0.68 \pm 0.16$&   1.81& 13.75& 1.21 & Y &         \\
JS 298  & $8\>39\>31.84$& $19\>24\>18.3$& 1.17&   229.8& S& 
  $ 1.68 \pm 0.17$&   4.47& 17.71& 1.59*& Y?&         \\
JS 305  & $8\>39\>38.56$& $19\>44\>19.5$& 2.31&    69.1& S& 
  $ 1.25 \pm 0.19$&   3.31& 17.92&      & Y?& $(c)$   \\
JS 329  & $8\>40\>00.78$& $19\>18\>35.5$& 0.86&   173.9& S& 
  $ 2.03 \pm 0.23$&   5.38& 14.43& 1.38 & Y &         \\
JS 349  & $8\>40\>15.35$& $19\>27\>31.7$& 3.11&    49.7& S& 
  $ 0.59 \pm 0.11$&   1.56& 13.91& 1.33 & Y &         \\
JS 706  & $8\>39\>15.15$& $19\>43\>31.6$& 0.42&   420.6& S& 
  $ 5.41 \pm 0.45$&  14.36& 16.96& 1.48*& Y &         \\
JS 718  & $8\>40\>04.26$& $19\>24\>51.0$& 1.16&   330.5& S& 
  $ 1.80 \pm 0.16$&   4.78& 18.02& 1.73*& Y &         \\
HSHJ 246& $8\>39\>03.31$& $19\>24\>06.3$& 9.95&    63.5& S& 
  $ 2.07 \pm 0.35$&   5.49&      & 1.70*& Y &         \\
HSHJ 258& $8\>39\>18.78$& $19\>22\>46.0$& 5.63&    10.4& P& 
  $ 2.31 \pm 0.76$&   1.33&      & 2.01*& Y & PhB     \\
HSHJ 283& $8\>39\>41.65$& $19\>29\>01.3$& 1.17&    30.0& S& 
  $ 0.34 \pm 0.08$&   0.89&      & 2.07*& Y & PhB     \\
HSHJ 289& $8\>39\>47.25$& $19\>39\>34.9$& 1.09&   132.5& S& 
  $ 0.88 \pm 0.11$&   2.32&      & 1.68*& Y &         \\
HSHJ 291& $8\>39\>54.52$& $19\>27\>37.9$& 1.51&    91.4& S& 
  $ 0.57 \pm 0.08$&   1.51&      & 1.77*& Y &         \\
HSHJ 293& $8\>40\>02.70$& $19\>40\>35.6$& 2.90&    48.3& S& 
  $ 0.58 \pm 0.11$&   1.53&      & 1.45*& Y &         \\
HSHJ 302& $8\>40\>11.62$& $19\>39\>12.7$& 0.79&   796.8& S& 
  $ 2.91 \pm 0.18$&   7.73&      & 1.45*& Y &         \\
HSHJ 328& $8\>40\>41.77$& $19\>30\>05.3$& 4.79&    32.5& S& 
  $ 0.61 \pm 0.13$&   1.62&      & 2.07*& Y & PhB     \\
WRS 4   & $8\>39\>12.62$& $19\>30\>14.4$& 2.82&   100.7& S& 
  $ 0.38 \>(g)   $&   0.88& 20.79&      & Y & $(d)$   \\
J0839531+192403& $8\>39\>53.23$& $19\>24\>04.0$& 0.41& 766.5& S& 
  $ 3.31 \pm 0.20$&   8.77&      &      & Y?& $(e)$   \\
J0840394+194255& $8\>40\>39.32$& $19\>42\>54.1$& 2.18& 35.3& M& 
  $ 2.06 \pm 0.49$&   5.47&      &      & Y & $(f)$   \\ 
\noalign{\smallskip}
\hline
\noalign{\smallskip}
\multicolumn{12}{l}{$(a)$ Luminosity in the {\it ROSAT} band ($0.1-2.4$ keV)
in units of $10^{28}$ erg~s$^{-1}$; \hskip 0.2cm $(b)$ An X indicates ROSAT
detections}\\ 
\multicolumn{12}{l}{$(c)$ $V-I_{\rm k} = 2.41$; \hskip 0.2cm $(d)$ $V-I_{\rm
k} = 3.46$; \hskip 0.2cm $(e)$ $J=13.94$, $K_{\rm s} = 13.03$; \hskip 0.2cm
$(f)$ $J=13.48$, $K_{\rm s} = 12.61$}\\
\multicolumn{12}{l}{$(g)$ The X-ray source is blended with GSC2 N231230081115:
the total count rate is $1.14\pm 0.13$, WRS 4 contributes by $\sim 1/3$ (see
text)}\\
\multicolumn{12}{l}{Star names are from \citet[KW]{kw27}, \citet[JC]{jc83},
\citet[JS]{js91}, \citet[HSHJ]{hshj95},}\\ 
\multicolumn{12}{l}{and \citet[WRS]{wrs95}. For the last two stars from
\citet{adams02} the 2MASS identification is given}\\
\end{tabular}
\label{detmemb}
\end{table*}

\begin{table*}
\caption{$3\sigma$ upper limits for undetected Praesepe members. Upper
limits have been derived from the summed dataset}
\begin{tabular}{lccrrrlll}
\hline\hline
\noalign{\smallskip}
Name& RA$_{\rm X}$& DEC$_{\rm X}$& count rate& $L_{\rm X}$& $V$& $B-V$& Mem&
Notes \\
 & \multicolumn{2}{c}{(J2000)}& (cts/ksec)& ($a$)& & 
$(R-I)_{\rm c}$\hskip -3pt *& & ($b$)\\
\hline
\noalign{\smallskip}
KW 150  & $8\>39\>06.11$& $19\>40\>36.5$& 0.54& 1.44&  7.45& 0.26 & Y?& \\
KW 229  & $8\>39\>57.78$& $19\>32\>29.2$& 0.19& 0.51&  7.54& 0.25 & Y & SB2,
X\\
KW 276  & $8\>40\>18.10$& $19\>31\>55.1$& 0.21& 0.55&  7.54& 0.16 & Y & SB \\
KW 284  & $8\>40\>20.14$& $19\>20\>56.4$& 0.40& 1.06&  6.77& 0.27 & Y & SB2\\
KW 300  & $8\>40\>27.01$& $19\>32\>41.3$& 0.22& 0.59&  6.30& 0.17 & Y?& SB2\\
KW 348  & $8\>40\>56.30$& $19\>34\>49.2$& 1.29& 3.43&  6.78& 0.17 & Y?& SB \\
KW 573  & $8\>40\>13.83$& $19\>44\>55.9$& 0.80& 2.11& 16.06& 1.50 & Y & \\
JC 212  & $8\>40\>31.55$& $19\>41\>43.2$& 0.41& 1.10& 15.20& 1.47 & Y & \\
JS 254  & $8\>38\>53.59$& $19\>34\>17.0$& 0.62& 1.63& 14.16& 1.33 & Y & \\
JS 301  & $8\>39\>36.51$& $19\>29\>07.9$& 0.25& 0.66& 15.12& 1.48 & Y & \\
JS 713  & $8\>39\>46.70$& $19\>44\>12.0$& 0.46& 1.23& 18.35&      & Y?&
$(V-I)_{\rm k} = 2.82$\\
HSHJ 304& $8\>40\>13.00$& $19\>45\>49.7$& 0.81& 2.14&      & 2.00*& Y & \\
RIZpr 11& $8\>39\>47.72$& $19\>28\>03.6$& 0.22& 0.59&      &      & Y &
$I_{\rm c} =19.47$, $I_{\rm c}-K_{\rm UKIRT} = 3.61$ \\
\noalign{\smallskip}
\hline
\noalign{\smallskip}
\multicolumn{9}{l}{$(a)$ Luminosity in the {\it ROSAT} band ($0.1-2.4$ keV)
in units of $10^{28}$ erg~s$^{-1}$}\\
\multicolumn{9}{l}{$(b)$ An X indicates ROSAT detections. For the other
notes see text} \\
\multicolumn{9}{l}{RIZpr 11 is from \citet{pinfield97}}
\end{tabular}
\label{upplim}
\end{table*}

\begin{table*}
\caption{X-ray sources with at least an optical counterpart which is a
cluster non-member or has no membership information. See Table~\ref{detmemb}
for the meaning of the ``set'' column}
\begin{tabular}{ccrcrlcl}
\hline\hline
\noalign{\smallskip}
RA$_{\rm X}$& DEC$_{\rm X}$& ML& set& count rate& Optical ID& $\Delta r$&
Notes\\
\multicolumn{2}{c}{(J2000)}& & & (cts/ksec)& & ($\;\arcsec\;$)& \\
\hline
\noalign{\smallskip}
$8\>38\>58.02$& $19\>33\>35.6$&  597.3& S& $ 5.66 \pm 0.41$& 
  GSC2 N23123008603 & 1.96& \\
$8\>39\>04.18$& $19\>23\>39.0$&   17.8& P& $ 4.53 \pm 1.26$& 
  KW 144            & 5.56& non-member \\
$8\>39\>07.02$& $19\>21\>51.2$& 5731.7& S& $39.25 \pm 1.20$& 
  4C 19.31          & 2.92& quasar\\
$8\>39\>10.39$& $19\>33\>29.2$&  225.8& S& $ 3.47 \pm 0.33$& 
  GSC2 N23123008580 & 2.17& \\
$8\>39\>10.86$& $19\>35\>12.7$&   75.6& S& $ 1.01 \pm 0.17$& 
  GSC2 N23123009122 & 3.21& \\
	      &               &       &  &                 & 
  GSC2 N23123009123 & 4.86& \\
$8\>39\>14.25$& $19\>33\>23.4$&   33.7& S& $ 0.53 \pm 0.11$& 
  KW 163            & 2.69& non-member \\
$8\>39\>16.09$& $19\>42\>49.0$&  561.4& S& $ 4.95 \pm 0.34$& 
  GSC2 N23122206641 & 1.06& \\
$8\>39\>16.17$& $19\>38\>16.5$&  119.3& S& $ 1.84 \pm 0.24$& 
  GSC2 N23122205633 & 7.87& \\
$8\>39\>26.39$& $19\>36\>58.9$&   32.0& S& $ 0.44 \pm 0.10$& 
  GSC2 N23122205462 & 4.34& \\
$8\>39\>36.65$& $19\>45\>10.8$&  357.9& S& $ 3.76 \pm 0.31$& 
  GSC2 N23122207129 & 0.81& \\
$8\>39\>43.41$& $19\>25\>13.3$&  429.1& S& $ 2.08 \pm 0.16$& 
  KW 206            & 0.69& non-member, $B-V=1.40$ \\
$8\>39\>50.97$& $19\>19\>12.5$&   15.1& S& $ 0.49 \pm 0.14$& 
  KW 211            & 6.99& non-member \\
$8\>39\>55.80$& $19\>36\>35.0$&   10.3& P& $ 0.97 \pm 0.33$& 
  GSC2 N23122205423 & 7.52& \\
$8\>39\>56.87$& $19\>43\>00.9$&  301.7& S& $ 2.13 \pm 0.19$& 
  WJJP 95           & 2.84& non-member, $B-V=0.90$, Wang et al.\\
$8\>40\>05.07$& $19\>38\>18.2$& 1398.9& S& $ 3.98 \pm 0.19$& 
  AKS 83            & 2.72& \citet{aks95} \\
$8\>40\>09.54$& $19\>31\>53.6$&   62.3& S& $ 0.44 \pm 0.08$& 
  KW 262            & 2.55& non-member \\
$8\>40\>10.73$& $19\>26\>31.9$&   39.3& S& $ 0.45 \pm 0.09$& 
  KW 266            & 3.27& non-member, $B-V=0.41$\\
$8\>40\>11.87$& $19\>28\>24.5$&   60.5& S& $ 0.55 \pm 0.09$& 
  GSC2 N23122203751 & 5.61& \\
$8\>40\>13.00$& $19\>28\>05.3$&  155.9& S& $ 0.97 \pm 0.11$& 
  GSC2 N23122203733 & 1.01& \\
              &               &       &  &                 & 
  GSC2 N23122203734 & 6.08& \\
$8\>40\>21.72$& $19\>21\>14.1$&   16.1& S& $ 0.41 \pm 0.12$& 
  2MASS J0840219+192117& 4.86& \\
$8\>40\>35.17$& $19\>32\>03.1$&  165.9& S& $ 1.16 \pm 0.13$& 
  KW 317            & 1.09& non-member, $B-V=0.58$\\
$8\>40\>37.97$& $19\>36\>33.2$&   12.8& S& $ 0.27 \pm 0.09$& 
  2MASS J0840379+193628& 4.57& \\ 
$8\>40\>39.04$& $19\>24\>44.7$&  279.3& S& $ 4.51 \pm 0.44$& 
  GSC2 N23122203337 & 8.42& \\
$8\>40\>47.30$& $19\>32\>39.6$&  183.4& S& $ 1.53 \pm 0.17$& 
  KW 333            & 2.59& non-member, $B-V=0.90$\\
\noalign{\smallskip}
\hline
\end{tabular}
\label{detnonm}
\end{table*}

Source detection was then performed on the combined dataset. The resulting
count rates are MOS equivalent count rates. We detected 183 sources with $ML
>10$ (corresponding to $4\sigma$) on the combined image; 16 additional
sources were detected above the same level only on the single instrument
images, giving a total number of 199 sources (note that for sources detected
only in the PN, the derived count rates are PN count rates).\footnote{
A problem was recently discovered in the ML detection code (XMM-{\it Newton}
News no. 29), which overestimates the ML values by a factor of $\sim 2$.
Only two of the sources with $ML<20$ are identified with cluster members
(KW~246 and HSHJ~258): a visual inspection of the images shows that KW~246
is a real source, while HSHJ~258 could be dubious. In any case, excluding
them from our analysis would not affect our results.}

In order to find optical counterparts to our source list, we have determined
the optimal search radius by constructing the cumulative distribution of the
offsets between X-ray and optical position, following RS95. We have adopted
a value of $6\arcsec$, which is expected to give less than 1 spurious
identification; however, we chose to use a larger value of $10\arcsec$ for
sources at off-axis angles greater than 10 arcmin, in order to take into
account possible errors in the X-ray positions due to the distorsion
introduced by the XMM-{\it Newton} optics. We find that, of the 199 X-ray
sources, 48 have a cluster member counterpart within the search radius,
including the two giants. For the remaining members with no associated X-ray
source we estimated 3$\sigma$ upper limits from the background count rates
at the optical position. The X-ray and optical properties of detected and
undetected cluster members are given in Tables~\ref{detmemb} and
\ref{upplim}, respectively. 

Nine additional sources have been identified with cluster non-members from
our input optical catalog; all of them have proper motion membership
probabilities $P_\mu =0$. We have found possible optical counterparts for
other 15 sources by cross-correlating the X-ray source list with the SIMBAD
database\footnote{operated at CDS, Strasbourg}, the Guide Star Catalog
II\footnote{The Guide Star Catalog II is a joint project of the Space
Telescope Science Institute and INAF/Osservatorio Astronomico di Torino} and
the 2MASS catalog\footnote{The 2MASS is a joint project of the University of
Massachusetts and the Infrared Processing and Analysis Center/California
Institute of Technology}: one of them is a known quasar, while for the
others no membership information is available. All these sources are listed
in Table~\ref{detnonm}. For the remaining 127 sources we did not find any
known counterpart in any of the available astronomical catalogs. 

\section{Results and discussion}
\label{results}

\subsection{X-ray luminosities}

In order to compare our results with previous observations, we have computed
X-ray fluxes for both detections and upper limits in the {\it ROSAT} band,
i.e. $0.1-2.4$ keV. A set of conversion factors (CFs) for MOS and PN were
computed using PIMMS assuming single-temperature Raymond-Smith plasmas with
temperatures in the range $\log T = 6.5-7.2$ K and $N_H = 3.8\times 10^{19}$
cm$^{-2}$, derived assuming an interstellar hydrogen volume density $n_{\rm
H} = 0.07$ cm$^{-3}$ \citep{paresce84} and the {\it Hipparcos} distance of
$180^{+10.7}_{-9.6}$ pc \citep{robich99}. For each temperature we then
compared the derived fluxes of common sources in the MOS and PN that are
associated with stellar counterparts, and selected the temperature giving
consistent values in the two cameras, i.e. such that the median ratio of the
PN to MOS fluxes was $\simeq 1$. This constraint is verified for $\log T
=6.8$; the corresponding adopted CF values are CF$ = 6.84 \times 10^{-12}$
erg~cm$^{-2}$~cnt$^{-1}$ for a single MOS camera and CF$ = 1.48 \times
10^{-12}$ erg~cm$^{-2}$~cnt$^{-1}$ for the PN camera. If we allow a 5\%
variation in the median PN/MOS ratio (i.e. $1\pm 0.05$), which corresponds
to $\Delta \log T = \pm 0.1$, the derived CFs differ by less than 5\%. Using
a series of two-temperature models with $\log T$ in the same range yields
similar results, with at most a 10\% difference in the derived CF values. We
note that the MOS CF is used for all sources detected on the summed dataset
or on the single MOS datasets, while for sources detected only on the PN the
PN CF is applied. X-ray luminosities are then computed using the above
mentioned {\it Hipparcos} distance.

The sensitivity reached in the central 10 arcmin of the field is $L_{\rm X}
\simeq 5.6 \times 10^{27}$ erg~s$^{-1}$, i.e. a factor $\sim 4$ higher than
that of the highest sensitivity region of the previous {\it ROSAT}
observation. The sensitivity is a factor of $\sim 2$ lower in the outer
parts of the image, and drops to $L_{\rm X} \simeq 2.1 \times 10^{28}$
erg~s$^{-1}$ in the region covered only by the MOS instruments.

\begin{figure}
\resizebox{\hsize}{!}{\includegraphics{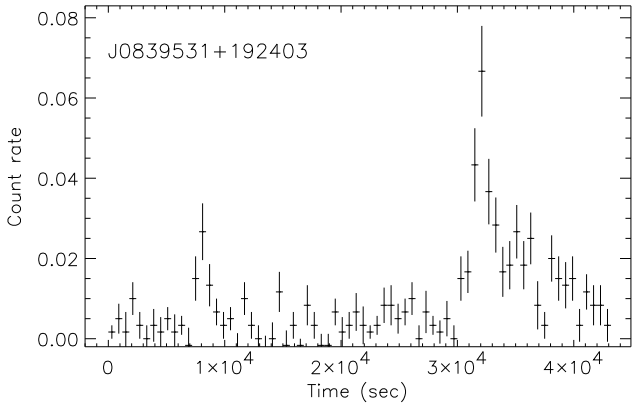}}
\caption{EPIC MOS1+MOS2+PN background-subtracted light curve of the M-type
Praesepe candidate 2MASS~J0839531+192403. Data are binned over 600 s}
\label{lcurve}
\end{figure}

\subsection{Detection rates of cluster members}
\label{detrate}

As mentioned in the previous section, 48 sources were identified with
cluster members. We detected \emph{all} the 17 F- and G-type stars included
in the XMM-{\it Newton} field of view, 10 of the 11 K-dwarfs (91\%) and 16
of the 22 M-dwarfs (73\%). These detection rates are considerably higher
than those obtained by RS95 in their {\it ROSAT} raster scan observation
($\sim 30$\% for F-G stars and $\sim 10$\% for K and M stars), thanks to the
higher sensitivity of our present XMM-{\it Newton} observation. Note however
that the detection rates of RS95 are considerably higher when considering
only Praesepe members included in our field of view; in particular, they
detected 7 of the 11 solar-type, i.e late F and G, stars (64\% -- one star,
KW\,287, was not included in their optical catalog).

One of the M-type Praesepe candidates, 2MASS J0839531+192403, underwent two
flares during our observation, with increases in the count rate by a factor
of $\sim 6$ for the first one and $\sim 15$ for the second one
(Fig.~\ref{lcurve}). The other sources show generally only low-level
variability below the $2\sigma$ level. 

\begin{figure}
\resizebox{\hsize}{!}{\includegraphics{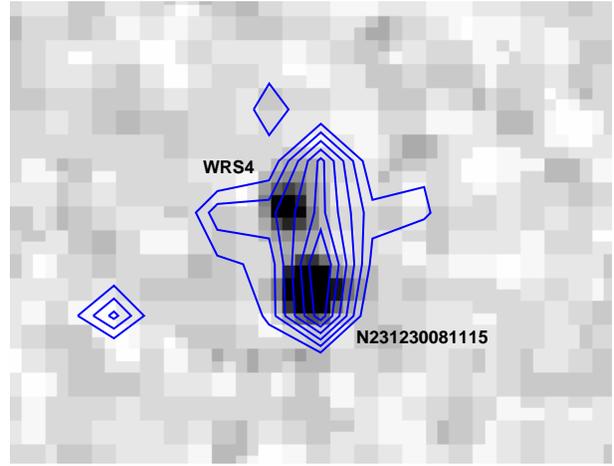}}
\caption{2MASS H-band image of WRS\,4 and GSC2\,N231230081115, with
superimposed the contour levels of the blended X-ray source. X-ray contours
correspond to 0.24, 0.32, 0.40, 0.48, 0.56, 0.64 cts/arcsec$^2$}
\label{wrs4}
\end{figure}

We finally mention the detection of the very low-mass star WRS\,4 which is
well below the fully convective boundary, having an estimated mass equal to
$0.13 \,M_{\sun}$ \citep{wrs95}. The X-ray source associated with WRS\,4 is
blended with the emission from the nearby (7$\arcsec$) star GSC2
N231230081115; the detection algorithm is not able to separate them (see
Fig.~\ref{wrs4}). We have estimated the contribution from WRS\,4 using two
approaches: (a) by measuring the counts in two circular regions centered on
the positions of the two stars, and (b) by fitting the distribution of
counts with two gaussian components; in both cases we find that WRS\,4
contributes by a factor $\sim 0.3$ to the total source count rate of $1.14
\pm 0.16$ cts/ksec. We therefore derive for WRS\,4 an X-ray luminosity
$L_{\rm X} \simeq 8.8 \times 10^{27}$ erg~s$^{-1}$, and an X-ray over
bolometric luminosity ratio $\log L_{\rm X}/L_{\rm bol}\simeq -3.1$. The
field of view contains also RIZpr\,11 which is one of the faintest known
Praesepe members, with an estimated mass $M \sim 0.09 \,M_{\sun}$, i.e. very
close to the substellar limit \citep{pinfield97,hodgkin99}; for this object
we obtain an upper limit $L_{\rm X} < 5.9 \times 10^{27}$ erg~s$^{-1}$
($\log L_{\rm X}/L_{\rm bol} < -2.7$). 

\begin{figure}
\resizebox{\hsize}{!}{\includegraphics{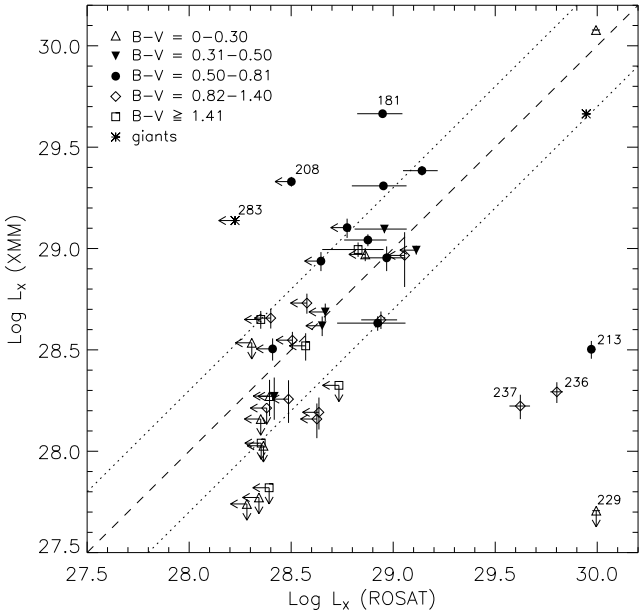}}
\caption{Comparison of the X-ray luminosities of Praesepe members falling in
the XMM-{\it Newton} field of view derived from XMM-{\it Newton} and {\it
ROSAT}. Different symbols are used for A stars (\emph{open triangles}),
early F stars (\emph{filled triangles}), late F-G stars (\emph{filled
circles}), K stars (\emph{diamonds}), M stars (\emph{squares}) and giants
(\emph{asterisks}). $1\sigma$ errors are shown for X-ray detections, while
upper limits are indicated by an arrow (where error bars are not shown, they
are smaller than the symbol size). Dotted lines indicate variations of a
factor of two in the X-ray luminosity. Stars that have changed by more than
this factor are labeled with their corresponding KW numbers
\protect\citep{kw27}}
\label{comp}
\end{figure}

We have detected both late-type giants (KW\,212 and KW\,283) falling in our
field of view. KW\,212 is the third brightest source, with $L_{\rm X}= 4.6
\times 10^{29}$ erg~s$^{-1}$, while KW\,283 is a factor of 3 weaker ($L_{\rm
X}= 1.4 \times 10^{29}$ erg~s$^{-1}$).

Three A-type stars ($B-V \le 0.3$) have been detected in our observation.
These stars do not possess strong massive winds and cannot generate magnetic
fields via the dynamo process due to the lack of a convection zone, thus
they should not be X-ray emitters. As discussed in several papers
\citep[e.g.][and references therein]{micela96}, the most likely possibility
is that their X-ray emission is due to an unseen late-type companion.
Indeed, two of them (KW\,224 and KW\,279) are known SB1 binaries. In
particular KW\,224 is the brightest source in our sample, with an X-ray
luminosity of $1.2 \times 10^{30}$ erg~s$^{-1}$. 

\subsection{Comparison with previous {\it ROSAT} observations}

\begin{figure*}
\resizebox{\hsize}{!}{\includegraphics[angle=90]{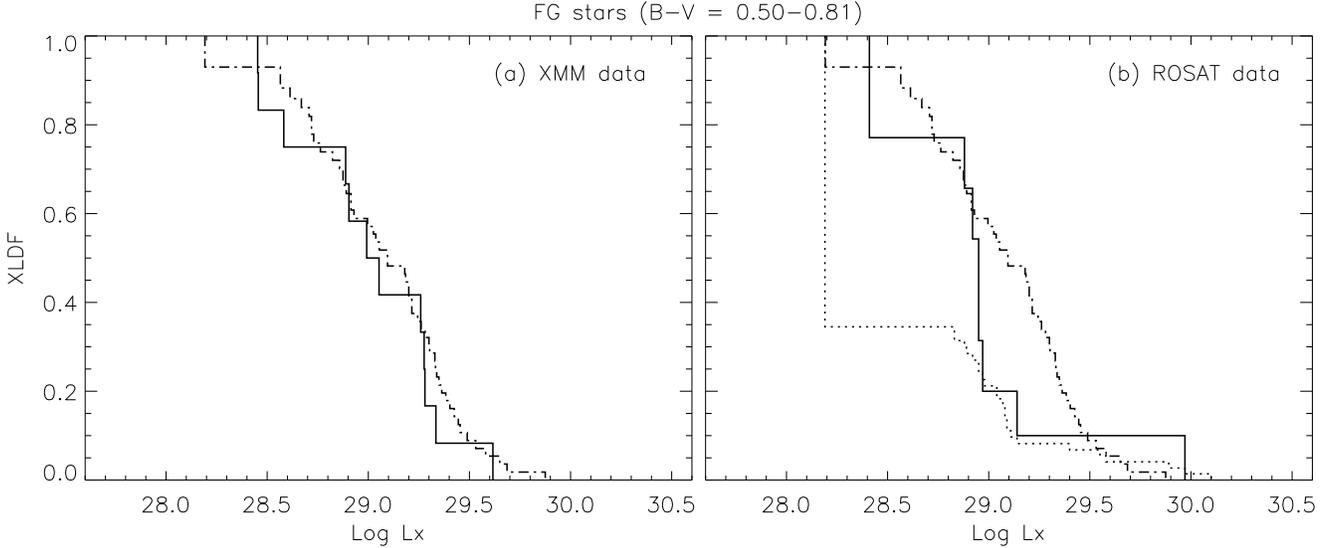}}
\caption{Comparison of the XLDF of Praesepe (solid line) and the Hyades
(dash-dotted line) for late F-G dwarf stars in the colour range $0.50 \le
B-V \le 0.81$. In panel (a) we show the Praesepe XLDF obtained from the
present XMM-{\it Newton} data. Panel (b) shows the XLDF derived, for the
same sample of stars, from the previous {\it ROSAT} observations; for
comparison, we also show the Praesepe XLDF derived from the entire {\it
ROSAT} sample (dotted line). The Hyades XLDF was derived from the {\it
ROSAT} observations by \protect\citet{stern95}}
\label{xldf_FG}
\end{figure*}

RS95 provide X-ray data for 44 members in common with our observation (the
other members missing from the {\it ROSAT} sample were not yet known or were
considered nonmembers at that time). The XMM-{\it Newton} and {\it ROSAT}
X-ray luminosities of the stars in common are compared in Fig.~\ref{comp}.
The {\it ROSAT} luminosities have been rescaled in order to account for the
new Hipparcos distance. We have checked that the CF used by RS95 is
consistent with our choice of the spectral model (i.e. $\log T=6.8$),
therefore no further correction is required in order to compare the results.

For most of the stars in common {\it ROSAT} luminosities are comparable with
those derived by us, with differences by at most a factor of two. Moreover,
for the stars detected only by XMM-{\it Newton} the upper limits inferred by
RS95 are generally comparable to our estimate of the X-ray luminosities.
There are however a few stars that show significant differences between the
two observations. In particular, four stars (KW\,213, KW\,229, KW\,236 and
KW\,237) have {\it ROSAT} luminosities more than an order of magnitude
higher than those derived by us. However, these stars are close together and
to KW\,224 and KW\,212, which are among the brightest sources in both
XMM-{\it Newton} and {\it ROSAT} (see the two points in the upper right
corner of Fig.~\ref{comp}). In the {\it ROSAT} image all these sources are
blended in a single source. A careful check of the {\it ROSAT} image shows
that the greatest contribution to the blended source comes indeed from
KW\,224 and KW\,212, with the other stars lying in the wings of the two
brightest ones; we therefore conclude that their flux had been largely
overestimated by RS95 because of confusion. 

Three stars were detected by us at higher luminosity levels than RS95
(KW\,181, KW\,208 and the giant KW\,283); in particular, KW\,208 and KW\,283
were not detected by RS95 with inferred upper limits lower than the
luminosity derived by us. The EPIC light curves of these stars show some
low-level variability but no evidence of flares. The difference in their
emission levels between {\it ROSAT} and XMM-{\it Newton} is therefore likely
due to long-term variability, possibly linked to activity cycles. We mention
that possible long-term variability was also detected among Hyades giants
\citep{stern95} based on the comparison between {\it ROSAT} and {\it
Einstein} luminosities.

\subsection{Comparison with the Hyades cluster}

In order to compare in a consistent way our results with the Hyades, we have
recomputed the CF for the {\it ROSAT} PSPC observations of the Hyades using
the same model spectrum with $\log T=6.8$. We find CF$ = 8\times 10^{-12}$
erg~cm$^{-2}$~cnt$^{-1}$, which is higher than the value ($6\times
10^{-12}$) adopted by \citet{stern95} and \citet{pye94}; X-ray luminosities
have been derived using the mean Hipparcos distance of $46.34\pm 0.27$ pc
\citep{perryman98}. Although the comparison between Praesepe and the Hyades
might be affected by uncertainties in the distance, spectral assumptions,
column density and the use of different instruments, a rough estimate of
these effects shows that their cumulative effect is small ($\sim 10-20$\%)
and does not significantly affect our results.

XLDFs for both clusters have been computed using the {\sc asurv} (Astronomy
SURvival Analysis) Ver. 1.2 software package \citep{feignels85,isobe86}.

\subsubsection{Solar-type stars}

In Fig.~\ref{xldf_FG}a we compare the XLDF of the Praesepe solar-type (i.e.
late F--G, $0.50 \le B-V \le 0.81$) members in our field of view with that
of the Hyades. As clearly shown in the figure, our data do not evidence the
discrepancy found by RS95, although the XLDF of Praesepe is still slightly
below that of the Hyades; the median luminosity ($\log L_{\rm X}=28.99$
erg~s$^{-1}$) as well as the 25th and 75th percentiles ($\log L_{\rm
X}=29.28$ and 28.58 erg~s$^{-1}$, respectively) are only slightly lower than
those of the Hyades ($\log L_{\rm X}=29.07$, 29.33 and 28.74 erg~s$^{-1}$,
respectively).

Our results for solar-type stars seem to contradict the previous results by
RS95 based on the {\it ROSAT} raster scan survey of Praesepe. However if one
considers only the subsample of Praesepe stars in the survey of RS95 in
common with the present sample, the disagreement between the XMM-{\it
Newton} and {\it ROSAT} results is considerably reduced
(Fig.~\ref{xldf_FG}b). In fact, as shown in the previous section, the 11
solar-type stars in common have generally {\it ROSAT} luminosities or upper
limits consistent with the X-ray luminosities derived by us. The difference
in the detection rates between the two observations is therefore due to the
different sensitivities of the two surveys. Moreover, the median luminosity
that we derive based on {\it ROSAT} luminosities or upper limits for the 11
stars in common ($\log L_{\rm X} = 28.93$ erg~s$^{-1}$) is comparable to the
median luminosity derived from the present XMM-{\it Newton} observations.

In other words, the overall discrepancy between the X-ray properties of the
Hyades and Praesepe pointed out by RS95 appears to be mostly due to X-ray
faint Praesepe members outside our XMM-{\it Newton} field of view (see
Fig.~\ref{xldf_FG}b). As a further check, we have compared the distribution
of {\it ROSAT} luminosities of stars falling inside and outside the XMM-{\it
Newton} field of view, performing a series of two-sample tests with {\sc
asurv}: the tests confirm that the two distributions are different, giving a
probability $\la 2$\% that they are drawn from the same population.

\begin{figure}
\resizebox{\hsize}{!}{\includegraphics[angle=-90,clip]{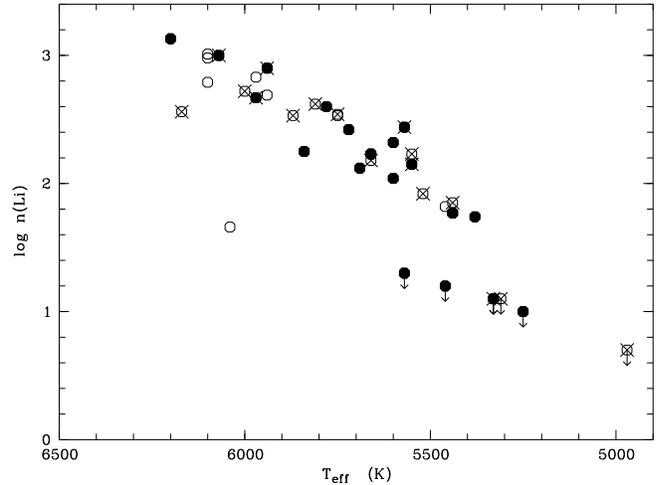}}
\caption{$\log n$(Li) vs. $T_{\rm eff}$ for Praesepe members with available
X-ray data. Lithium data have been retrieved from \protect\citet{sod93}.
Filled symbols denote stars with $\log L_{\rm X} \leq 28.6$ erg~s$^{-1}$,
while crossed symbols indicate photometric and/or spectroscopic binaries}
\label{Li}
\end{figure}

\begin{figure*}
\resizebox{\hsize}{!}{\includegraphics[angle=90]{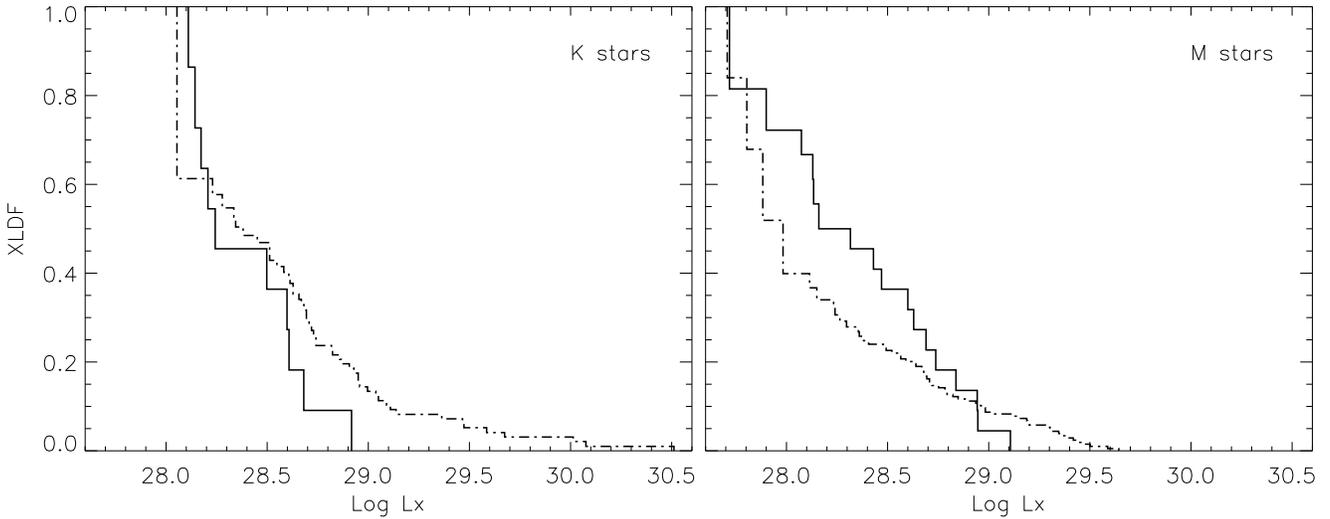}}
\caption{Comparison of the XLDF of Praesepe (solid line) and the Hyades
(dash-dotted line) for K dwarfs ($0.82 \le B-V \le 1.40$, \emph{left panel})
and M dwarfs ($B-V \ge 1.41$, \emph{right panel}). The Hyades XLDF was
derived from the {\it ROSAT} observations by \protect\citet{stern95} and
\protect\citet{pye94}} 
\label{xldf_KM}
\end{figure*}

As mentioned in the introduction, \citet{holland00} made the hypothesis that
Praesepe is formed by two merged clusters, and found that the X-ray
brightest sources are almost exclusively located in the ``main'' cluster.
Based on this fact, they suggested that an age spread may be present between
the cluster core and the subcluster stars. Our XMM-{\it Newton} observation
indeed covers a small area of the main cluster and thus presumably includes
the X-ray brightest (and possibly youngest) cluster population. 

In order to investigate whether Praesepe could be characterized by an age
dispersion, we compared the results on X--ray emission with Li abundances
derived by \citet{sod93}. These authors noted that Praesepe showed a larger
scatter in Li abundances than the Hyades and that confirmed cluster members
were present with a Li content well below the average trend. The direct
comparison of the spectra of stars with the same color, but different Li
abundances, allowed them to exclude that the scatter could be due to
measurement uncertainties. They also excluded that the presence of binary
stars in their sample could be the reason for the dispersion. 40 Praesepe
members in the range $0.5 \leq B-V \leq 0.81$ have both Li and X-ray
measurements available. In Fig.~\ref{Li} we show lithium abundances ($\log
n$(Li)) as a function of effective temperature for Praesepe members with
X-ray data available. Filled symbols denote stars with X-ray luminosity (or
upper limit) below the 25th percentile, or $\log L_{\rm X} \leq 28.6$
erg~s$^{-1}$. Note that, for stars with both ROSAT and XMM data available,
we used the latter. Crossed symbols indicate photometric and/or
spectroscopic binaries; updated information on binarity was retrieved from
\citet{barrado98}. The figure first shows that, as mentioned by
\citet{sod93}, binarity is not the main reason for the dispersion in Li.
Second, on one hand, not all the X-ray faint stars are lithium-poor; on the
other hand, all but two of the lithium-poor stars (defined as those with a
Li abundance more than a factor of 2 below the mean trend) have an X-ray
luminosity (or upper limit) below the median value. Whereas these arguments
are not conclusive, Li abundances could indeed support the hypothesis of
Praesepe being composed by two different subclusters.

\subsubsection{Late-type stars}

Fig.~\ref{xldf_KM} shows the comparison between the XLDF of Praesepe and the
Hyades for K dwarfs ($0.82 \le B-V \le 1.40$) and M dwarfs ($B-V \ge 1.41$).
The sample of K stars in Praesepe appears to be slightly less luminous than
the Hyades: the median luminosity is $\log L_{\rm X} = 28.22$ erg~s$^{-1}$
compared to 28.35 erg~s$^{-1}$ for the Hyades. On the contrary, Praesepe M
stars appear to be significantly more luminous, with a median $\log L_{\rm
X} = 28.16$ erg~s$^{-1}$, while the median luminosity of the Hyades M-dwarfs
is 27.90 erg~s$^{-1}$. The two-sample tests give a probability of $\le 10$\%
that the two distributions are drawn from the same population. This result
is in agreement with the results of \citet{barrado98}, who found that M
dwarfs in Praesepe were characterized by a higher chromospheric activity
than their Hyades counterparts. We have performed the same tests also for
the K dwarf samples, but the results are inconclusive ($p \sim 37-78$\%). We
also note that the XLDFs of Praesepe K and M stars lack the high luminosity
tail ($\log L_{\rm X} \geq 29.0$ erg~s$^{-1}$) which is instead evident in
the XLDFs of the Hyades. The high luminosity tail of the Hyades XLDFs is
mostly due to binaries, which have been shown to be significantly more
luminous than single stars of the same spectral type \citep{pye94}. Our
sample contains only a small fraction of late-type binaries, all of which
have been detected: however, their X-ray luminosity is comparable to that of
single stars.

\begin{table}
\caption{Best-fit parameters for KW\,224, KW\,212 and KW\,181. Errors are
90\% confidence ranges for one interesting parameter}
\begin{tabular}{lccc}
\hline\hline
\noalign{\smallskip}
  & KW\,224& KW\,212& KW\,181 \\
\hline
\noalign{\smallskip}
$kT_1$ (keV)                & $0.41_{-0.03}^{+0.02}$& $0.51_{-0.03}^{+0.03}$&
$0.39_{-0.06}^{+0.07}$\\[2pt]
$kT_2$ (keV)                & $0.91_{-0.04}^{+0.04}$& \ldots                &
$0.89_{-0.16}^{+0.14}$\\[2pt]
$EM_1$ ($10^{52}$~cm$^{-3}$)& $4.85_{-1.24}^{+0.58}$& $2.83_{-0.98}^{+0.50}$&
$3.47_{-1.00}^{+0.95}$\\[2pt]
$EM_2$ ($10^{52}$~cm$^{-3}$)& $4.81_{-0.70}^{+0.73}$& \ldots                &
$1.77_{-0.51}^{+1.26}$\\[2pt]
$Z/Z_{\sun}$                & $0.41_{-0.07}^{+0.08}$& $0.49_{-0.09}^{+0.30}$&
$0.23_{-0.07}^{+0.12}$\\[2pt]
$\chi^2_{\rm r}/d.o.f.$     & $1.11/196$            & $0.90/78$             &
$1.02/58$             \\[2pt]
$F_{\rm X}\,(10^{-13}$ erg~cm$^{-2}$~s$^{-1}$)& $3.5$& $1.2$                &
$1.3$\\
\noalign{\smallskip}
\hline
\noalign{\smallskip}
\multicolumn{4}{l}{$F_{\rm X}$ is the unabsorbed flux in the $0.1-2.4$ keV
band}\\
\end{tabular}
\label{fits}
\end{table}

\begin{figure}
\resizebox{\hsize}{!}{\includegraphics{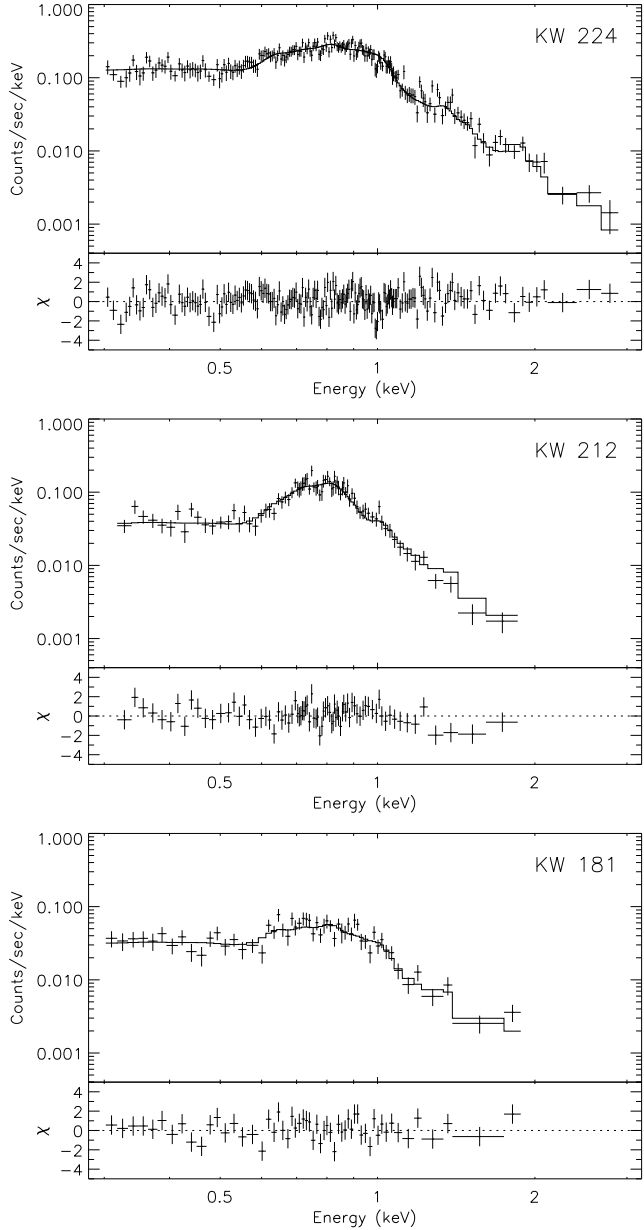}}
\caption{PN spectra of the three brightest sources in the Praesepe field.
The best-fit model is also shown. For KW\,212 and KW\,181 the emission is
dominated by background above 2 keV}
\label{spectra}
\end{figure}

\subsection{Spectral analysis of the brightest sources}

The three brightest Praesepe members in our sample (KW\,224, KW\,212 and
KW\,181) have more than 1000 counts in the PN, allowing us to perform a
spectral analysis of their emission. Spectra have been extracted using an
extraction region of radius $40\arcsec$ for KW\,224 and KW\,181, and
$25\arcsec$ for KW\,212 to avoid a nearby source. Background spectra were
extracted from equal circular regions free from X-ray sources and close to
each of the stars. Response matrices were generated for each source using
the standard SAS tasks. Spectra have been rebinned in order to have at least
20 counts per bin, and were fitted in XSPEC v.11.2.0 in the energy range
$0.3-2.5$ keV, using a two-temperature APEC model with varying global metal
abundance $Z$. In the case of KW\,212 we find that a single temperature
model is sufficient to fit the spectrum. The column density was kept fixed
to $N_{\rm H} = 3.8\times 10^{19}$ cm$^{-2}$, since its value was not
constrained by the fit.

The best-fit parameters are given in Table~\ref{fits}; the three spectra
together with the best-fit models are shown in Fig.~\ref{spectra}. We find
that the coronae of KW\,224 and KW\,181 have similar 2-T components of $\sim
5$ and $\sim 10$ MK, although the emission measure ratio is different
($EM_2/EM_1 \sim 1$ for KW\,224 and $\sim 0.5$ for KW\,181). The giant
KW\,212 is cooler, with a single temperature of $\sim 6$ MK. We find
subsolar abundances for all sources ($Z/Z_{\sun} \simeq 0.2-0.5$) in spite
of the almost solar photospheric metallicity of Praesepe stars. Subsolar
coronal abundances are commonly found in most stellar coronae, and in some
cases they are also significantly lower than the photospheric abundance,
e.g. in the young star AB~Dor \citep{maggio00,gudel01ab}. We mention that
similar results (with Fe $\sim 0.3-0.6$ Fe$_{\sun}$) have been found also
for two Hyades stars (the giant $\theta^1$~Tau and the solar-type star
VB~50) observed with XMM-{\it Newton} \citep{francio02mssl}.

The X-ray flux derived from the spectral fits is consistent with that
derived from the spatial analysis, within the 10\% uncertainty in the
assumed values of the CFs.

\section{Conclusions}
\label{concl}

In this paper we have presented the results of a new observation of the
Praesepe cluster performed with XMM-{\it Newton}, which resulted in the
detection of 100\% of the solar-type stars, 91\% of the K stars and 73\% of
the M dwarfs in the field of view. In particular, we have clearly detected
the very low-mass star WRS\,4, which is well below the fully convective
boundary, with $L_{\rm X}\sim 8.8 \times 10^{27}$ erg s$^{-1}$.

Our XMM-{\it Newton} observation shows that the X-ray properties of our
sample of Praesepe solar-type stars are in good agreement with the Hyades,
apparently in contrast with the previous {\it ROSAT} results; however, we
have shown that the discrepancy between Praesepe and the Hyades previously
found by {\it ROSAT} apparently shows up only when considering a much larger
area of the Praesepe cluster than the one covered by the XMM-{\it Newton}
field of view. This finding seems to support the suggestion made by
\citet{holland00} that Prasepe may be formed by two merged clusters of
different ages: our XMM-{\it Newton} pointing covers in fact a part of the
``main'' cluster, where the brightest X-ray sources are located. We mention,
however, that the large proper motion study by \citet{adams02} does not
suggest a peculiar kinematic history for Praesepe. New XMM-{\it Newton}
observations pointed in the outer regions of Praesepe are therefore required
to verify whether this hypothesis is correct.

Spectral analysis of the three brightest sources has shown that the coronae
of these stars have metallicities ranging from 0.2 to 0.5 times the solar
value, i.e. significantly lower than the measured photospheric abundance,
which is almost solar, in agreement with other observations of stellar
coronae. 

Most of the sources do not show evidence of variability greater than a
factor of 2 over a timescale of $\sim 10$ years. Only three sources are
significantly brighter during our XMM-{\it Newton} observation, likely as a
consequence of activity cycles. On the other hand, the sources with much
higher X-ray luminosity in {\it ROSAT} compared to the present observation
had been overestimated due to confusion.

Thanks to the higher sensitivity of XMM-{\it Newton} with respect to
previous satellites, we have detected a large number of new X-ray sources
with no known cataloged optical counterpart. Optical follow-up photometric
and spectroscopic observations will be required to ascertain their nature
and determine whether part of them are unknown faint members of the cluster.

\begin{acknowledgements}

We thank the anonymous referee for his useful comments and suggestions which
helped us to improve the paper. This research has made use of the Open
Cluster Database, provided by C.F. Prosser (deceased) and J.R. Stauffer,
currently available at http://cfa-www.harvard.edu/\~{}stauffer/opencl/ or by
anonymous ftp at cfa-ftp.harvard.edu (cd /pub/stauffer/clusters), and of the
WEBDA Database, provided by J.-C. Mermilliod, available at
http://obswww.unige.ch/webda/. We acknowledge partial support by Ministero
dell'Istruzione, Universit\`a e Ricerca (MIUR) and by the Italian Space
Agency (ASI). 

\end{acknowledgements}


\begin{thebibliography}{36}
\expandafter\ifx\csname natexlab\endcsname\relax\def\natexlab#1{#1}\fi

\bibitem[{Adams {et~al.}(2002)Adams, Stauffer, Skrutskie, Monet,
  Portegies~Zwart, Janes, \& Beichman}]{adams02}
Adams, J.~D., Stauffer, J.~R., Skrutskie, M.~F., {et~al.} 2002, AJ, 124, 1570

\bibitem[{Andruk {et~al.}(1995)Andruk, Kharchenko, Schilbach, \&
  Scholz}]{aks95}
Andruk, V., Kharchenko, N., Schilbach, E., \& Scholz, R.~D. 1995, AN, 316, 225

\bibitem[{Barrado~y Navascu{\`e}s {et~al.}(1998)Barrado~y Navascu{\`e}s,
  Stauffer, \& Randich}]{barrado98}
Barrado~y Navascu{\`e}s, D., Stauffer, J.~R., \& Randich, S. 1998, ApJ, 506,
  347

\bibitem[{Feigelson \& Nelson(1985)}]{feignels85}
Feigelson, E.~D. \& Nelson, P.~I. 1985, ApJ, 293, 192

\bibitem[{Franciosini {et~al.}(2000{\natexlab{a}})Franciosini, Randich, \&
  Pallavicini}]{francio00}
Franciosini, E., Randich, S., \& Pallavicini, R. 2000{\natexlab{a}}, A\&A, 357,
  139

\bibitem[{Franciosini {et~al.}(2000{\natexlab{b}})Franciosini, Randich, \&
  Pallavicini}]{francio00asp}
Franciosini, E., Randich, S., \& Pallavicini, R. 2000{\natexlab{b}}, in Stellar
  Clusters and Associations: Convection, Rotation, and Dynamos, ed.
  R.~Pallavicini, G.~Micela, \& S.~Sciortino, ASP Conf. Ser. No. 198 (San
  Francisco: ASP), 447

\bibitem[{Franciosini {et~al.}(2002)Franciosini, Sanz-Forcada, Maggio, \&
  Pallavicini}]{francio02mssl}
Franciosini, E., Sanz-Forcada, J., Maggio, A., \& Pallavicini, R. 2002, in MSSL
  Workshop on High resolution X-ray spectroscopy with XMM-Newton and Chandra,
  http://www.mssl.ucl.ac.uk/{\~{}}gbr/rgs\_workshop/papers/franci-osini\_e.ps

\bibitem[{Friel \& Boesgaard(1992)}]{friel92}
Friel, E.~D. \& Boesgaard, A.~M. 1992, ApJ, 387, 170

\bibitem[{G{\"u}del {et~al.}(2001)G{\"u}del, Audard, Briggs, Haberl, Magee,
  {et~al.}}]{gudel01ab}
G{\"u}del, M., Audard, M., Briggs, K., {et~al.} 2001, A\&A, 365, L336

\bibitem[{Hambly {et~al.}(1995)Hambly, Steele, Hawkins, \& Jameson}]{hshj95}
Hambly, N.~C., Steele, I.~A., Hawkins, M.~R.~S., \& Jameson, R.~F. 1995, A\&AS,
  109, 29

\bibitem[{Harmer {et~al.}(2001)Harmer, Jeffries, Totten, \& Pye}]{harmer01}
Harmer, S., Jeffries, R.~D., Totten, E.~J., \& Pye, J.~P. 2001, MNRAS, 324, 473

\bibitem[{Hodgkin {et~al.}(1999)Hodgkin, Pinfield, Jameson, Steele, Cossburn,
  \& Hambly}]{hodgkin99}
Hodgkin, S.~T., Pinfield, D.~J., Jameson, R.~F., {et~al.} 1999, MNRAS, 310, 87

\bibitem[{Holland {et~al.}(2000)Holland, Jameson, Hodgkin, Davies, \&
  Pinfield}]{holland00}
Holland, K., Jameson, R.~F., Hodgkin, S., Davies, M.~B., \& Pinfield, D. 2000,
  MNRAS, 319, 956

\bibitem[{Isobe {et~al.}(1986)Isobe, Feigelson, \& Nelson}]{isobe86}
Isobe, T., Feigelson, E.~D., \& Nelson, P.~I. 1986, ApJ, 306, 490

\bibitem[{Jeffries(1999)}]{jeffries99}
Jeffries, R.~D. 1999, in Solar and Stellar Activity: Similarities and
  Differences, ed. C.~J. Butler \& J.~G. Doyle, ASP Conf. Ser. No. 158 (San
  Francisco: ASP), 75

\bibitem[{Jones \& Cudworth(1983)}]{jc83}
Jones, B.~F. \& Cudworth, K. 1983, AJ, 88, 215

\bibitem[{Jones \& Stauffer(1991)}]{js91}
Jones, B.~F. \& Stauffer, J.~R. 1991, AJ, 102, 1080

\bibitem[{Klein-Wassink(1927)}]{kw27}
Klein-Wassink, W.~J. 1927, Publ. Kapteyn Astron. Lab. Groningen, 41, 1

\bibitem[{Maggio {et~al.}(2000)Maggio, Pallavicini, Reale, \&
  Tagliaferri}]{maggio00}
Maggio, A., Pallavicini, R., Reale, F., \& Tagliaferri, G. 2000, A\&A, 356, 627

\bibitem[{Mermilliod(1997)}]{merm97ms}
Mermilliod, J.-C. 1997, Mem. Soc. Astr. It., 68, 853

\bibitem[{Mermilliod {et~al.}(1990)Mermilliod, Weis, Duquennoy, \&
  Mayor}]{merm90}
Mermilliod, J.-C., Weis, E.~W., Duquennoy, A., \& Mayor, M. 1990, A\&A, 235,
  114

\bibitem[{Micela {et~al.}(1996)Micela, Sciortino, Kashyap, Harnden, \&
  Rosner}]{micela96}
Micela, G., Sciortino, S., Kashyap, V., Harnden, Jr., F.~R., \& Rosner, R.
  1996, ApJS, 102, 75

\bibitem[{Paresce(1984)}]{paresce84}
Paresce, F. 1984, AJ, 89, 1022

\bibitem[{Perryman {et~al.}(1998)Perryman, Brown, Lebreton, G{\'o}mez, Turon,
  {et~al.}}]{perryman98}
Perryman, M.~A.~C., Brown, A.~G.~A., Lebreton, Y., {et~al.} 1998, A\&A, 331, 81

\bibitem[{Pinfield {et~al.}(1997)Pinfield, Hodgkin, Jameson, Cossburn, \& von
  Hippel}]{pinfield97}
Pinfield, D.~J., Hodgkin, S.~T., Jameson, R.~F., Cossburn, M.~R., \& von
  Hippel, T. 1997, MNRAS, 287, 180

\bibitem[{Pizzolato {et~al.}(2001)Pizzolato, Ventura, D'Antona, Maggio, Micela,
  \& Sciortino}]{pizzol01}
Pizzolato, N., Ventura, P., D'Antona, F., {et~al.} 2001, A\&A, 373, 597

\bibitem[{Pye {et~al.}(1994)Pye, Hodgkin, Stern, \& Stauffer}]{pye94}
Pye, J.~P., Hodgkin, S.~T., Stern, R.~A., \& Stauffer, J.~R. 1994, MNRAS, 266,
  798

\bibitem[{Randich(2000)}]{randich00}
Randich, S. 2000, in Stellar Clusters and Associations: Convection, Rotation,
  and Dynamos, ed. R.~Pallavicini, G.~Micela, \& S.~Sciortino, ASP Conf. Ser.
  No. 198 (San Francisco: ASP), 401

\bibitem[{Randich \& Schmitt(1995)}]{randsch95}
Randich, S. \& Schmitt, J.~H.~M.~M. 1995, A\&A, 298, 115

\bibitem[{Robichon {et~al.}(1999)Robichon, Arenou, Mermilliod, \&
  Turon}]{robich99}
Robichon, N., Arenou, F., Mermilliod, J.-C., \& Turon, C. 1999, A\&A, 345, 471

\bibitem[{Sciortino {et~al.}(2000)Sciortino, Micela, Favata, Spagna, \&
  Lattanzi}]{sciorti00}
Sciortino, S., Micela, G., Favata, F., Spagna, A., \& Lattanzi, M.~G. 2000,
  A\&A, 357, 460

\bibitem[{Soderblom {et~al.}(1993)Soderblom, Fedele, Jones, Stauffer, \&
  Prosser}]{sod93}
Soderblom, D.~R., Fedele, S.~B., Jones, B.~F., Stauffer, J.~R., \& Prosser,
  C.~F. 1993, AJ, 106, 1080

\bibitem[{Stern {et~al.}(1995)Stern, Schmitt, \& Kahabka}]{stern95}
Stern, R.~A., Schmitt, J.~H.~M.~M., \& Kahabka, P.~T. 1995, ApJ, 448, 683

\bibitem[{Totten {et~al.}(2000)Totten, Jeffries, Harmer, \& Pye}]{totten00}
Totten, E., Jeffries, R., Harmer, S., \& Pye, J. 2000, in Stellar Clusters and
  Associations: Convection, Rotation, and Dynamos, ed. R.~Pallavicini,
  G.~Micela, \& S.~Sciortino, ASP Conf. Ser. No. 198 (San Francisco: ASP), 451

\bibitem[{Wang {et~al.}(1995)Wang, Chen, Zhao, \& Jiang}]{wang95}
Wang, J.~J., Chen, L., Zhao, J.~H., \& Jiang, P.~F. 1995, A\&AS, 113, 419

\bibitem[{Williams {et~al.}(1995)Williams, Rieke, \& Stauffer}]{wrs95}
Williams, D.~M., Rieke, G.~H., \& Stauffer, J.~R. 1995, ApJ, 445, 359

\end{thebibliography}

\end{document}